Article

# Preoperative Prediction of Catheter Ablation Outcome in Persistent Atrial Fibrillation Patients through Spectral Organization Analysis of the Surface Fibrillatory Waves

Pilar Escribano [1],*, Juan Ródenas [1], Manuel García [1], Miguel A. Arias [2], Víctor M. Hidalgo [3], Sofía Calero [3], José J. Rieta [4] and Raúl Alcaraz [1]

1 Research Group in Electronic, Biomedical and Telecommunication Engineering, University of Castilla-La Mancha, 02071 Albacete, Spain
2 Cardiac Arrhythmia Department, Complejo Hospitalario Universitario de Toledo, 45007 Toledo, Spain
3 Cardiac Arrhythmia Department, Complejo Hospitalario Universitario de Albacete, 02006 Albacete, Spain
4 BioMIT.org, Electronic Engineering Department, Universitat Politecnica de Valencia, 46022 Valencia, Spain
* Correspondence: pilar.escribano@uclm.es

**Abstract:** Catheter ablation (CA) is a commonly used treatment for persistent atrial fibrillation (AF). Since its medium/long-term success rate remains limited, preoperative prediction of its outcome is gaining clinical interest to optimally select candidates for the procedure. Among predictors based on the surface electrocardiogram, the dominant frequency (DF) and harmonic exponential decay ($\gamma$) of the fibrillatory waves ($f$-waves) have reported promising but clinically insufficient results. Hence, the main goal of this work was to conduct a broader analysis of the $f$-wave harmonic spectral structure to improve CA outcome prediction through several entropy-based measures computed on different frequency bands. On a database of 151 persistent AF patients under radio-frequency CA and a follow-up of 9 months, the newly introduced parameters discriminated between patients who relapsed to AF and those who maintained SR at about 70%, which was statistically superior to the DF and approximately similar to $\gamma$. They also provided complementary information to $\gamma$ through different combinations in multivariate models based on lineal discriminant analysis and report classification performance improvement of about 5%. These results suggest that the presence of larger harmonics and a proportionally smaller DF peak is associated with a decreased probability of AF recurrence after CA.

## 1. Introduction

Atrial fibrillation (AF) is the most frequently encountered cardiac arrhythmia in clinical practice [1]. This disruption of heart sinus rhythm (SR) affects roughly 37.5 million people worldwide [2], thus making it one of the most important public health problems and a significant cause of rising healthcare costs in developed countries [3]. Although the mechanisms underlying AF are not fully understood, it requires a combination of triggers, mainly located near the pulmonary veins (PVs), and a vulnerable atrial substrate characterized by reentry circuits [4]. AF is commonly linked to life-altering symptoms such as palpitations, fatigue, chest pain, shortness of breath, and dizziness [5]. This arrhythmia is not a direct cause of death, but it is associated with a two-fold increase in mortality because of a higher risk of heart failure and ischemic stroke [6].

Depending on the duration and recurrence of arrhythmic episodes, AF is classified into several stages [7]. However, the disease is not static and often progresses from paroxysmal to sustained forms over the course of a few years; this is usually correlated with irreversible electrical and structural remodeling of the atrial substrate that, in turn, promotes perpetuation of the arrhythmia [4]. Hence, early diagnosis of AF and finding the best way to restore SR as soon as possible is highly advisable [8]. To that purpose, pharmacological or electrical cardioversion strategies are frequently used in clinical practice until they are no longer effective or not recommended. In that case, catheter ablation (CA) has become a commonly used alternative [9] thanks to its superior ability over antiarrhythmic drugs for restoring SR in the midterm [10]. This therapy has evolved during the last two decades, and the most common protocol remains based on PVs isolation (PVI) [11]. Despite its usually successful initial outcome, it is not as effective in patients with persistent AF compared to patients with paroxysmal AF in the medium/long term, since approximately 40% to 50% of them relapse to AF within the first year [9]. Thus, the high complexity in the management of catheters and the complications associated with the procedure encourage careful assessment of the benefits and risks for each patient [12].

This tailored assessment, along with the clinical, social, and economic challenges that AF will involve in the coming decades [2], motivate the emerging clinical interest in the development of preoperative predictors of CA outcome, which could be helpful in the selection of patients who would benefit the most from the procedure [13]. Some benefits of choosing optimal candidates for CA are the reduction of hospitalization rates, limitation of repeated procedures, assignment of more appropriate AF treatments to improve patient quality of life, and minimization of the risks and costs associated with AF treatment [13]. So far, some clinical indices have been proposed to anticipate midterm CA outcome, such as the total duration of AF, the duration of the last arrhythmic episode, left atrial diameter, and history of hypertension or diabetes, among others. However, most of these parameters have reported controversial results, so there is no strong evidence to consider them as reliable single predictors of AF recurrence after the procedure [14].

Some authors have pioneered the use of parameters manually measured on the surface electrocardiogram (ECG) obtained prior to CA. Uncoordinated electrical conduction in the atria means the typical P-wave is replaced on the ECG by undulatory activity known as fibrillatory waves ($f$-waves). Manual measurement of the time between these waves, i.e., the AF cycle length (AFCL) [15], and their amplitude [16,17] have been positively correlated with midterm CA results. During ablation, invasive measures of the AFCL or its inverse, i.e., the dominant frequency (DF), in different atrial structures have also been confirmed to have a direct link with AF recurrence several months after the procedure [18–21]. However, these indices have the disadvantage of being based on manual and/or invasive ECG metrics, which entails a high degree of subjectivity in measurement during or after CA intervention.

To overcome these limitations and to achieve predictors available for the selection of candidates for CA, more recent studies have addressed non-invasive extraction of the aforementioned metrics through automated processing of the preoperative surface ECG. Researchers have reported performance at least as good as previously proposed clinical parameters and their manually or invasively derived versions [22]. Beyond the single DF and $f$-wave amplitude (FWA), the presence of large harmonics of the DF has also been identified as a promising predictor of CA outcome [18,23]. Higher harmonic content in the $f$-wave spectral distribution has been associated with a greater degree of organization of the atrial electrical activity and with higher probability of maintaining SR after different AF treatments, including CA [18,23], electrical cardioversion [24,25], and pharmacological therapy [26]. However, only the power contained by the DF and its harmonics has been analyzed to date, and the main goal of the present work is, hence, to conduct a broader analysis of the $f$-wave harmonic spectral structure to improve preoperative prediction of CA outcome in persistent AF patients.

## 2. Materials and Methods

### 2.1. Study Population

The population enrolled in this work consisted of 151 AF patients (35 women and 116 men) between 20 and 82 years old, with 59 years being the rounded mean age of the group. They were consecutively treated with radiofrequency CA for the first time following standard clinical indications at two Spanish hospitals (i.e., University Hospitals of Toledo and Albacete), thus constituting a retrospective database. The ablation procedure started with patient sedation using general anesthesia or conscious sedation after suspending all antiarrhythmic drug therapy except amiodarone >5 half-lives before the intervention. Anticoagulant drugs were also used to avoid thromboembolic complications. Indeed, an initial bolus of heparin was administered, followed by additional doses properly activated by coagulation-time monitoring throughout the procedure. Isolation of PVs was achieved by creating electrically impenetrable boundaries surrounding their ostia using a radiofrequency source [27]. Thus, a catheter was used to generate point-by-point lesions through the release of radiofrequency current for at least 30 s and then creating a contiguous antral circumferential line around the PVs, whose location were determined using a mapping catheter [28]. The procedure finished when all PVs were successfully isolated. If AF still remained at that point, SR was restored by electrical cardioversion.

The procedure was initially successful in all patients, who were monitored for several hours after the intervention without presenting any important complications. They received anticoagulant and antiarrhythmic drugs according to clinical judgment and had a quite standard follow-up with a visit, ECG, and 24 h Holter at 3 and 9 months, as recommended by the current 2020 ESC guidelines [7]. Thereafter, visits each 12 months were planned. A blanking period of 3 months was considered to define every episode of AF or other arrhythmias lasting more than 30 s as recurrence, but the patients were advised to go at any moment to an emergency room in case of AF-related symptoms.

*2.2. Data Preprocessing*

The database of this study consisted of a standard 12-lead ECG signal continuously recorded just before starting the CA procedure while the patient was under AF. Hence, a total of 151 ECG recordings with variable durations of between approximately 6 s and 5 min were available. The signals were acquired by the equipment available in the hospitals, with 16-bit resolution and a sampling rate of 977 Hz. For the present study, lead V1 was selected since it contains the $f$-waves with the greatest amplitude regarding ventricular activity, thus favoring their automated extraction and accurate analysis. In fact, the electrode that records this unipolar lead is located in a standard position very close to the atria, so it is the most appropriate lead for recording AF activity [29].

The selected ECG signal was then preprocessed to minimize the perturbations unavoidably recorded together with the electrical heart activity and, consequently, to improve further analysis of the $f$-waves. Since the patient was at rest during recording of the surface ECG before CA, the main disturbances presented on the ECG consisted of baseline wander, powerline interference, and disturbances associated with high-frequency noise sources [30]. The low-frequency component associated with baseline wander was estimated using a low-pass filter with a cut-off frequency of 0.8 Hz and zero-phase distortion through an IIR structure and forward/backward filtering and then subtracted from the original ECG signal [30]. Powerline interference was removed by a denoising algorithm based on the stationary wavelet transform. It applies a novel threshold function to the wavelet coefficients for filtering the unwanted frequency and its harmonics, simultaneously minimizing the distortion of the rest of the components [31]. The algorithm was also able to remove most of the high-frequency noise, but a low-pass forward/backward IIR filter with a cut-off frequency of 70 Hz was additionally used to make the ECG signal as clean as possible [30].

Finally, $f$-wave characterization requires that ventricular activity is first canceled. Although there are different techniques for this, $f$-waves were automatically extracted from the preprocessed ECG signal using a well-established QRST cancellation technique based on adaptive singular value cancellation (ASVC). The algorithm firstly obtained a cancellation template from the singular-value decomposition of a set of QRST complexes that were temporally aligned with regard to the R-peak. Then, the resulting template was adapted in amplitude for the cancellation of each single QRST complex. Notably, this method avoids spikes caused by discontinuities between QRST complexes and the subsequent TQ intervals through a softening approach that considers the differences between the cancellation template and the QRST segment at its beginning and end to minimize sudden transitions [32]. As an example, Figure 1 shows the $f$-waves obtained from a typical preprocessed ECG interval.

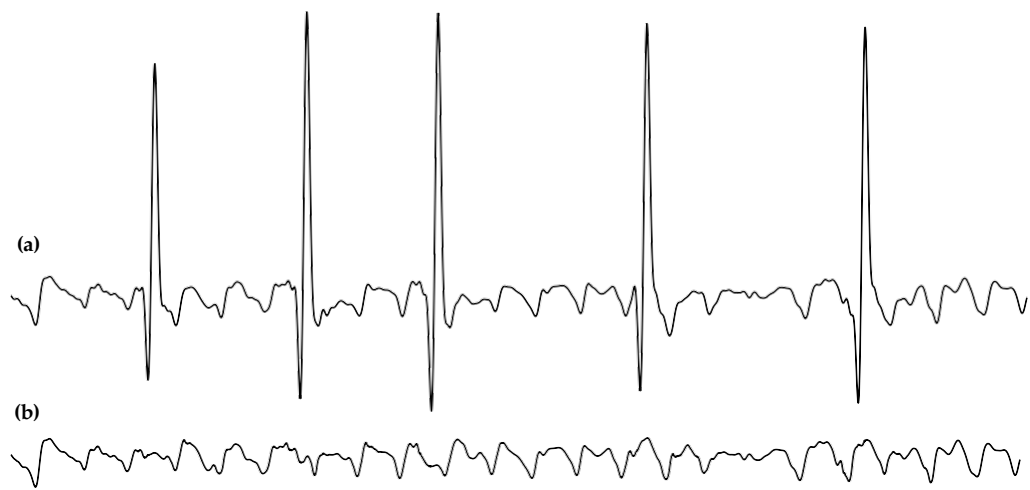

**Figure 1.** Example of a common (**a**) preprocessed ECG interval and (**b**) its $f$-waves extracted via a well-known QRST cancellation method [32].

*2.3. Spectral Characterization of the $f$-Waves*

Given the disparity in the ECG length acquired from the patients, the preprocessed recordings were segmented into 6 s intervals. For each patient, five consecutive intervals at most were considered, and the ECG-derived parameters from each one were averaged for unbiased subject-based analysis and classification. In this way, comparable values were obtained for all patients regardless of ECG signal duration, and slight variability noticed in $f$-waves owing to extracardiac noise [33] was minimized.

The power spectral density (PSD) of each 6 s ECG excerpt was estimated using the Welch Periodogram and is referred to as $W(f)$. The computational parameters of this algorithm were selected to provide a spectral resolution of 0.1 Hz with a Hamming window 4000 points long and with 3000 points overlapping between adjacent windowed sections [23]. To serve as a reference, common spectral features previously proposed to anticipate CA outcome were then automatically computed from the $f$-wave segments. Thus, the DF was obtained as the frequency with the highest PSD amplitude [34,35], with this power value also considered in the study. Both parameters hereafter are referred to as $f_0$ and $W(f_0)$, respectively. Making use of a 1 Hz bandwidth window centered on $2 \cdot f_0$, the first harmonic of the DF ($f_1$) and its spectral power ($W(f_1)$) were also computed. The harmonic exponential decay ($\gamma$) was additionally estimated as a measure of the presence of harmonic components of the DF [25,26]. This parameter was defined as the logarithmic ratio of the spectral power between the DF and its first harmonic, i.e.

$$\gamma = \ln\left(\frac{W(f_0)}{W(f_1)}\right). \tag{1}$$

The power of the DF and its harmonics has also been typically quantified through the well-known organization index ($O$). It was defined as the ratio of the cumulative power under the DF and its first two harmonics and the area of the entire power spectrum between 3 and 25 Hz [36]. Both parameters $\gamma$ and $O$ consider only the power under the largest frequency components in the spectral distribution of the $f$-waves, not the shape and morphology of the harmonics, i.e., how the power is distributed around each frequency peak. Hence, in the present work, some pioneering parameters were considered for broader spec-tral characterization of the $f$-waves. More precisely, the shape of the DF and its harmonic components were separately analyzed by dividing the $f$-wave spectral distribution into two bands, as Figure 2 shows. The low-frequency (LF) band considered the spectral content around the DF, whereas the high-frequency (HF) band accounted for harmonic content. The cut-off frequency separating both bands was positioned approximately halfway between the DF and its first harmonic, that is, at three means of the DF value ($f_0$). For a global overview of the spectral distribution of the $f$-waves, the total frequency (TF) band ranging from 3 to 25 Hz was also analyzed.

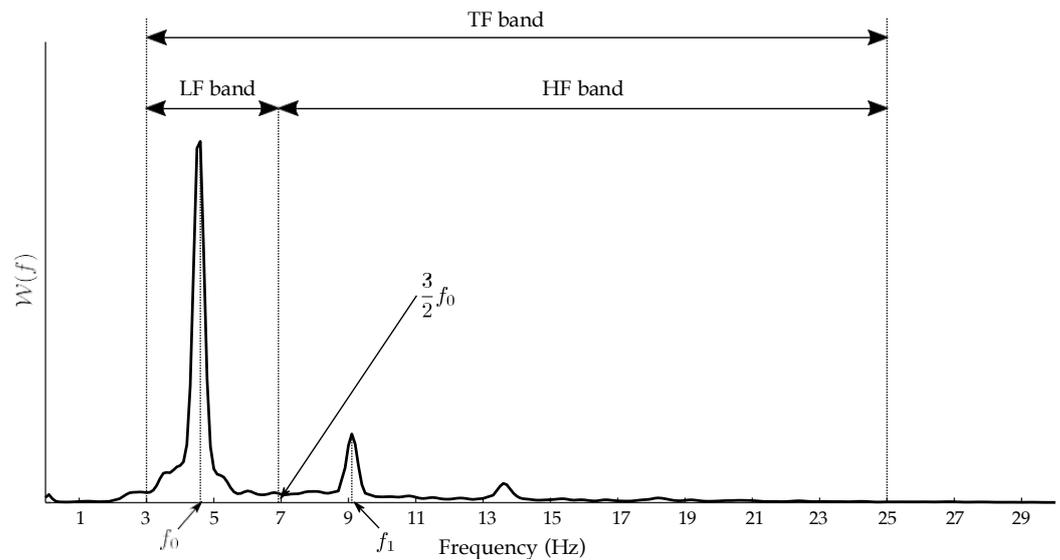

**Figure 2.** Spectral distribution of a common excerpt of $f$-waves divided into two frequency bands, i.e., low-frequency (LF) band and high-frequency (HF) band. Both form the total frequency (TF) band, ranging between 3 and 25 Hz.

The power distribution within each one of the three frequency bands was assessed through several entropy-based measures. Particularly, Wiener entropy, also known as spectral-flatness measure ($F$), was firstly computed by dividing geometric and arithmetic means of the power contained by each band [37,38]; that is

$$\mathcal{F} = \frac{\sqrt[N]{\prod_{f=f_l}^{f_u} \mathcal{W}(f)}}{\frac{1}{N} \sum_{f=f_l}^{f_u} \mathcal{W}(f)}, \qquad (2)$$

where $f_l$ and $f_u$ are the lower and upper frequency band limits, respectively, and $N$ is the total number of frequency samples. This index estimates the uniformity of signal-energy distribution in the frequency domain, so high values indicate more uniform or flat distributions, and low values are more peaky ones [38]. Another measure providing similar information is spectral entropy ($S$), but this index was defined in a completely different way. In fact, $S$ quantifies spectral complexity of a signal by computing the sparseness of its spectral distribution via Shannon entropy [39,40]. In brief, the PSD of the signal has to first be normalized by the total power of the frequency band of interest to obtain a probability function with unit area, i.e.,

$$\widetilde{\mathcal{W}}(f) = \frac{\mathcal{W}(f)}{\sum_{f=f_l}^{f_u} \mathcal{W}(f)}. \qquad (3)$$

Then, $S$ is estimated by computing the Shannon entropy from the resulting probability function, so that

$$S = -\frac{1}{\ln(N)} \sum_{f=f_l}^{f_u} \widetilde{\mathcal{W}}(f) \cdot \ln\left(\widetilde{\mathcal{W}}(f)\right). \quad (4)$$

It should be noted that $S$ is normalized by the highest possible value, i.e., $\ln(N)$, and thus ranges between 0 and 1. A high value of $S$ implies a flat, uniform spectrum with a broad spectral content, whereas a low value implies a spectrum with all the power condensed into a single frequency bin, i.e., a less complex, more predictable signal [41].

A generalized version of $S$ can be obtained by replacing the Shannon entropy with the Rényi entropy, thus obtaining Rényi spectral entropy ($R$) as

$$\mathcal{R} = \frac{1}{\ln(N)} \cdot \frac{1}{1-\alpha} \ln\left(\sum_{f=f_l}^{f_u} \widetilde{\mathcal{W}}(f)^\alpha\right), \quad (5)$$

where $\alpha$ ($\alpha \geq 0$ and $\alpha \neq 1$) is a bias parameter. Note that the Shannon entropy is an instance of Rényi entropy for $\alpha = 1$, and therefore, $R$ might be a more sensitive tool than $S$ to detect subtle changes in spectral distribution of a time series if $\alpha$ is appropriately chosen [41].

Finally, another measure of the spectral complexity of a signal is $C_0$ complexity. This index is more robust to conditions of nonlinearity and non-stationarity than previous ones [42]. For its computation, the normalized PSD of the $f$-waves was modified to only preserve its most irregular part, i.e.,

$$\widehat{\mathcal{W}}(f) = \begin{cases} \widetilde{\mathcal{W}}(f), & \text{if } \widetilde{\mathcal{W}}(f) \leq T, \\ 0, & \text{if } \widetilde{\mathcal{W}}(f) > T, \end{cases} \quad (6)$$

with $T$ being a threshold computed from the mean spectral power, so that

$$T = \frac{2}{N} \sum_{f=f_l}^{f_u} \widetilde{\mathcal{W}}(f). \quad (7)$$

Then, the $C_0$ complexity is computed as the power ratio of the irregular part to the original signal [42], i.e.,

$$C_0 = \sum_{f=f_l}^{f_u} \frac{\widehat{\mathcal{W}}(f)}{\widetilde{\mathcal{W}}(f)}, \quad (8)$$

resulting in a real number between 0 and 1, so that the greater the predominance of the irregular part of the signal, the higher the value of $C_0$.

*2.4. Statistical Analysis and Classification Performance*

Normality and homoscedasticity of continuous variables were firstly assessed using Lilliefors and Levene's tests, respectively. When both conditions were met, a parametric Student's *t*-test was used to measure statistical differences between the two groups of patients, i.e., those who maintained SR and those who relapsed to AF after a follow-up of 9 months. A non-parametric Mann–Whitney U-test was used for the same purpose when data distributions were non-normal but homoscedastic. Since most parameters showed normal and homoscedastic distributions, values of all the features are summarized along the manuscript in terms of mean ± standard deviation. Regarding categorical variables, they are reported as number and percentage, and were compared using a Fisher exact test. In all cases, a value of significance $p < 0.05$ was considered as statistically significant.

On the other hand, the classification performance of each single feature was evaluated through a repeated, patient-based, 10-fold cross-validation approach [43]. Precisely, in each cross-validation procedure, the data were first partitioned into 10 equally sized folds. Subsequently, 10 iterations of training and validation were performed, such that within each one a different fold of the data was held out for validation while the remaining nine folds were used for learning. The data were stratified by ensuring that each fold was a good representative of the whole. Linear discriminant analysis (LDA) was used to train a prediction model in each iteration. This analysis assumes that different classes generate data based on Gaussian distributions, such that training an LDA model involves finding the parameters for a Gaussian distribution for each class. In fact, the procedure searches for a projection hyperplane of the observations for which the variance of each class is minimized and the distance between the means of the classes is maximized [44].

The classification results for each 10-fold cross-validation procedure are summarized by means of a receiver operating characteristic (ROC) curve computed on the classification scores provided by the obtained LDA models. This plot is the result of plotting the fraction of true positives out of total positives (sensitivity) against the fraction of false positives out of total negatives (1-specificity) at various thresholds. Sensitivity (Se) was considered to be the rate of patients who were correctly classified as relapsing to AF, whereas specificity (Sp) was the percentage of patients properly identified as maintaining SR. An optimal threshold for separating both groups was selected to provide the best balance between Se and Sp, although in this way, the highest percentage of patients correctly classified, i.e., accuracy (Acc), could not be achieved [45]. The area under the ROC curve (AUC) was also obtained as an aggregate performance measure across all possible thresholds [45]. To provide additional information about the proportions of positive and negative samples that were true positives and true negatives, the positive predictive value (PPV) and negative predictive value (NPV) were also computed. The described validation process was repeated 100 times to obtain general and unbiased classification outcomes [43]. The data were reshuffled and re-stratified before each 10-fold cross-validation approach, and Se, Sp, Acc, AUC, PPV, and NPV values were averaged for the 100 cycles.

Finally, to explore complementary information among single features and to improve prediction of CA outcome after a follow-up of 9 months, a multivariate analysis was conducted. In this case, a LDA was also used to build prediction models based on linear combinations of those features automatically selected by making use of a forward sequential selection technique. In this analysis, features were sequentially added to an empty candidate set until the addition of further features did not decrease the criterion function, i.e., the prediction error [46]. That error was assessed inside repeated cross-validation loops to avoid any bias in feature selection [43]. Hence, variables that were selected more frequently for the models were used last to build several LDA-based prediction algorithms and were evaluated as single features, i.e., by running the patient-based 10-fold cross-validation 100 times. The classification improvement achieved by these models regarding single features was statistically evaluated using an asymptotic McNemar's test. To compare the accuracies of two classification models, the algorithm first compared their predicted labels against the true labels and then detected whether the difference between the misclassification rates was statistically significant [47].

## 3. Results

After a follow-up of 9 months, 103 patients maintained SR, and the remaining 48 relapsed to AF. This implies that the CA procedure was unsuccessful in the mid-term for 31.79% of the patients, which is consistent with current AF recurrence statistics [9,11]. The baseline clinical characteristics of both groups are provided in Table 1. As can be seen, none of the features collected for the patients, i.e., gender, age, AF duration before the CA procedure, body mass index, and left atrium diameter, presented statistically significant differences between the patients who maintained SR and those who relapsed to AF.

**Table 1.** Baseline clinical characteristics of the population enrolled in the study with their corresponding statistical significance to distinguish between the two groups of patients.

| Clinical Feature | Rhythm after Follow-Up | | *p*-Value |
| --- | --- | --- | --- |
|  | SR | AF |  |
| Number of patients (%) | 103 (68.21%) | 48 (31.79%) | — |
| Male (%) | 79 (76.70%) | 37 (77.08%) | 1.000 |
| Age (years) | 59.37 ± 12.24 | 57.23 ± 12.82 | 0.326 |
| With AF <1 year (%) | 6 (5.83%) | 6 (12.50%) | 0.198 |
| With AF 1–3 years (%) | 70 (67.96%) | 31 (64.58%) | 0.713 |
| With AF >3 years (%) | 27 (26.21%) | 11 (22.92%) | 0.841 |
| Body mass index (kg/m$^2$) | 27.79 ± 3.55 | 29.06 ± 4.86 | 0.073 |
| Left atrium diameter (mm) | 44.11 ± 5.70 | 45.65 ± 5.32 | 0.117 |

Regarding the wide range of features considered in the present work to characterize *f*-wave spectral distribution, Table 2 shows the values obtained for the two group of patients and the corresponding statistical significance (*p*-value). Among the parameters commonly proposed in previous works, $f_0$, $f_1$, and $\gamma$ provided statistically significant differences between patients who maintained SR and relapsed to AF, with the former having lower values for the three indices. Similarly, the four proposed entropy- and complexity-based features also provided statistically significant differences between both groups of patients when the TF band covering the whole *f*-wave distribution between 3 and 25 Hz was considered. In the case of the LF band, mainly containing the DF component, *F* and $C_0$ also provided statistically significant differences between the two groups of patients, and *R* reported a tendency close to being significant. On the contrary, no relevant differences between the groups of patients were noticed in any of the four indices in the frequency band covering the harmonic content, i.e., the HF band. It should be noted that several values of $\alpha$ between 0.1 and 2 were tested to compute the Rényi spectral entropy, but the best results (i.e., those presented in Table 2) were obtained for $\alpha = 0.1$.

The classification performance of the single features reporting statistically significant values of *p* < 0.05 is displayed in Table 3. Most indices presented values of Se, Sp, Acc, and AUC greater than 60%, but only $\gamma$ and $C_{0TF}$ exhibited performance metrics higher than 70%. Both parameters reported similar classification performances, which were statistically better than that provided by $f_0$ for all the conducted validation cycles according to McNemar's test. These two variables also provided the highest values of PPV and NPV, about 52% and 83%, thus improving the DF results by more than 15 and 10%, respectively. Similarly, the classification results reported by Wiener entropy and Rényi spectral entropy for the TF band (i.e., $F_{TF}$ and $R_{TF}$) were also statistically superior to that of $f_0$, but in this case, values of Acc and AUC of about 67%, values of PPV of about 47%, and values of NPV of about 81% were obtained. Contrarily, no statistically significant differences in the classification performance was noticed between $f_0$ and the entropy-based indices computed on the LF band, i.e., $F_{LF}$ and $C_{0LF}$.

On the other hand, multivariate analysis showed that the parameters most frequently selected for the LDA-based models built throughout the validation cycles were $\gamma$ and $F_{TF}$. Nonetheless, instead of the last parameter, $C_{0TF}$ and $R_{TF}$ were sometimes selected, along with the index $\gamma$. The classification performance of these three LDA-based models is presented in the first rows of Table 4. As can be observed, they were very similar for the three cases, reaching values of Acc and AUC of about 75% and values of PPV and NPV of about 58% and 86%, respectively. Notably, the three prediction models obtained classification results statistically better than those of the included single features (i.e., $\gamma$, $F_{TF}$, $C_{0TF}$, and $R_{TF}$) and better than $f_0$ for most validation cycles according to McNemar's test. In fact, improvements in values of Acc, AUC, PPV, and NPV greater than 5% were obtained by the LDA-based models compared to those of the included single features and by about 20% compared to those of the DF. Finally, the inclusion of a third variable to these prediction models did not improve the classification results. For instance, the last three rows of Table 4 display how Acc and AUC slightly decreased when the models were complemented with another feature, even when it was related to the DF component (such as $R_{LF}$) and was sometimes chosen by the automated feature selection algorithm. Along a similar line, the inclusion of any clinical and/or echocardiographic variables presented in Table 1 improved classification performance of the prediction models.

**Table 2.** Mean and standard deviation values for the analyzed metrics from the two groups of patients and their corresponding statistical significance (*p*-value).

| Feature | Rhythm after the Follow-Up | | *p*-Value |
|---|---|---|---|
| | SR | AF | |
| $f_0$ (Hz) | 5.69 ± 1.12 | 6.14 ± 0.99 | 0.009 |
| $W(f_0)$ (mV$^2$) | 0.00051 ± 0.00091 | 0.00060 ± 0.00092 | 0.059 |
| $f_1$ (Hz) | 11.34 ± 2.25 | 12.28 ± 1.98 | 0.008 |
| $W(f_1)$ (mV$^2$) | 0.00013 ± 0.00041 | 0.00005 ± 0.00009 | 0.754 |
| $\gamma$ | 2.20 ± 0.77 | 2.80 ± 0.57 | <0.001 |
| $O$ | 0.52 ± 0.15 | 0.53 ± 0.12 | 0.488 |
| $F_{LF}$ | 0.48 ± 0.15 | 0.43 ± 0.12 | 0.017 |
| $S_{LF}$ | 0.82 ± 0.08 | 0.81 ± 0.06 | 0.224 |
| $R_{LF}$ | 0.980 ± 0.010 | 0.978 ± 0.008 | 0.065 |
| $C_{0LF}$ | 0.45 ± 0.14 | 0.41 ± 0.11 | 0.045 |
| $F_{HF}$ | 0.49 ± 0.13 | 0.50 ± 0.10 | 0.865 |
| $S_{HF}$ | 0.86 ± 0.07 | 0.87 ± 0.05 | 0.828 |
| $R_{HF}$ | 0.985 ± 0.007 | 0.986 ± 0.005 | 0.967 |
| $C_{0HF}$ | 0.48 ± 0.10 | 0.49 ± 0.08 | 0.550 |
| $F_{TF}$ | 0.28 ± 0.10 | 0.22 ± 0.06 | <0.001 |
| $S_{TF}$ | 0.76 ± 0.07 | 0.73 ± 0.06 | 0.008 |
| $R_{TF}$ | 0.974 ± 0.008 | 0.971 ± 0.006 | <0.001 |
| $C_{0TF}$ | 0.32 ± 0.07 | 0.28 ± 0.05 | <0.001 |

**Table 3.** Classification between patients who relapsed to AF and maintained SR during the follow-up based on the most-predictive single parameters.

| Feature | Se (%) | Sp (%) | Acc (%) | AUC | PPV (%) | NPV (%) |
|---|---|---|---|---|---|---|
| $f_0$ | 56.25 | 56.42 | 56.36 | 0.617 | 37.56 | 73.45 |
| $f_1$ | 56.33 | 56.68 | 56.57 | 0.620 | 37.73 | 73.58 |
| $\gamma$ | 69.29 | 69.18 | 69.22 | 0.728 | 51.17 | 82.86 |
| $F_{LF}$ | 60.27 | 60.33 | 60.31 | 0.606 | 41.45 | 76.52 |
| $C_{0LF}$ | 56.54 | 56.71 | 56.66 | 0.587 | 37.84 | 73.68 |
| $F_{TF}$ | 66.13 | 66.22 | 66.19 | 0.676 | 47.71 | 80.75 |
| $S_{TF}$ | 58.83 | 59.28 | 59.14 | 0.623 | 40.24 | 75.55 |
| $R_{TF}$ | 66.10 | 66.18 | 66.16 | 0.675 | 47.67 | 80.73 |
| $C_{0TF}$ | 70.83 | 70.91 | 70.89 | 0.703 | 53.16 | 83.92 |

**Table 4.** Classification between patients who relapsed to AF and maintained SR during the follow-up provided by the prediction models obtained from multivariable linear discriminant analysis.

| Features in the Model | Se (%) | Sp (%) | Acc (%) | AUC | PPV (%) | NPV (%) |
|---|---|---|---|---|---|---|
| $\gamma$ and $F_{TF}$ | 72.75 | 75.14 | 74.38 | 0.758 | 57.69 | 85.54 |
| $\gamma$ and $R_{TF}$ | 72.77 | 74.89 | 74.22 | 0.748 | 57.46 | 85.51 |
| $\gamma$ and $C_{0TF}$ | 73.04 | 74.66 | 74.15 | 0.759 | 57.33 | 85.60 |
| $\gamma$, $F_{TF}$, and $C_{0TF}$ | 72.15 | 74.56 | 73.79 | 0.751 | 56.93 | 85.17 |
| $\gamma$, $F_{LF}$, and $R_{TF}$ | 73.90 | 73.12 | 73.36 | 0.743 | 56.16 | 85.74 |
| $\gamma$, $R_{TF}$, and $C_{0TF}$ | 70.31 | 74.90 | 73.44 | 0.753 | 56.63 | 84.41 |

## 4. Discussion

Despite its limited success rate in the medium/long term, CA remains a commonly used treatment for persistent AF [9]. Hence, clinical interest in the preoperative prediction of the outcome of this intervention has substantially grown in the last years [1], since it might anticipate those patients with a high probability of early AF recurrence [48]. Keeping in mind that CA often involves long-lasting procedures and high costs and risks for patients unable to maintain SR for prolonged periods of time, the selection of optimal candidates for this treatment would have interesting benefits. In this respect, not only could the huge costs associated with AF treatment be reduced, but the risks related to the CA procedure would be limited, tailored approaches could be enabled, and the number of repeated interventions could be minimized in many patients, among other benefits [13].

So far, some clinical predictors of arrhythmia recurrence after CA have been explored, such as AF duration, anatomical characteristics of the left atrium, presence of concomitant diseases, etc. However, they have only provided controversial results with limited predictive ability, which is usually lower in persistent rather than paroxysmal AF patients [14,49,50]. Accordingly, in the present study, none of the evaluated clinical variables were relevant in the prediction of patients who relapsed to AF versus those who maintained SR after a follow-up of 9 months. On the other hand, several previous works have introduced the development of predictors based on quantitative analysis of atrial electrical conduction. Invasive studies have shown that the DF and its inverse (i.e., the AFCL) have a moderate predictive capacity of midterm SR maintenance after CA [18–21,51,52]. However, these predictors require invasive recordings of the atrial electrical activity by means of electrograms acquired during the procedure, thus entailing an unnecessary risk in those patients for whom the CA procedure will not be successful.

Subsequently, non-invasive studies have also proven the predictive potential of manual measurement of the AFCL on surface ECGs, yielding a higher preoperative value for patients with higher probability of long-term SR maintenance after CA [15]. More recently, once $f$-waves were extracted from the surface ECG, the automated measurement of the DF has also shown a reasonable ability to anticipate CA outcome in persistent AF patients, associating lower values of frequency with a lower probability of AF recurrence in the medium/long term [34,35]. The results obtained by the DF in the present work are consistent with that tendency (see Table 2), and since this index has been directly linked to the degree of electrical remodeling presented by the atria [34], it can be considered a good reference for comparison with other predictors. The frequency of the DF's first harmonic ($f_1$) also presented the same trend in the results for both groups of patients, exhibiting predictive ability similar to that of the DF with Acc, AUC, PPV, and NPV values about 57%, 62%, 38%, and 74%, respectively (see Table 2).

In contrast to these results, the power peak both for the DF and its first harmonic did not reveal statistically significant differences between patients who relapsed to AF and those who maintained SR after the follow-up. Along the same line, the organization index $O$ was also unable to discern between the two groups of patients, thus suggesting that the power globally concentrated around the DF and its first harmonic is not relevant in the prediction of CA outcome. Similar results have also been reported in other works [23,36]. However, the power ratio of the first harmonic to the DF, i.e., the index $γ$, proved to be one of the best single predictors, with values of Acc and AUC of about 70%, and values of PPV and NPV of about 51% and 83%, respectively (see Table 2). A similar finding was previously outlined in a previous work, where $γ$ was the only spectral parameter able to report a statistically relevant significance on a limited dataset of 22 persistent AF patients under radio-frequency CA [23]. In the present work, that outcome has been corroborated on a much wider database of 151 patients. According to the definition of $γ$, low values indicate the presence of strong harmonics of the DF, and the obtained results hence suggest that the higher the harmonic content of the DF, the lower the probability of AF recurrence after CA. This finding is confirmed in Figure 3, which shows mean spectral distributions for all the 6 s ECG segments from the patients who relapsed to AF compared to those who maintained

SR after the follow-up once they were aligned with respect to the DF. As can be seen, on average, the patients who maintained SR exhibited greater harmonics of the DF than those who relapsed to AF. Interestingly, the presence of larger harmonics has also been associated with increased probability of AF termination during CA [18] and with medium/long-term maintenance of SR after diverse AF therapies, including pharmacological cardioversion [26] and internal [24] and external electrical cardioversion [25].

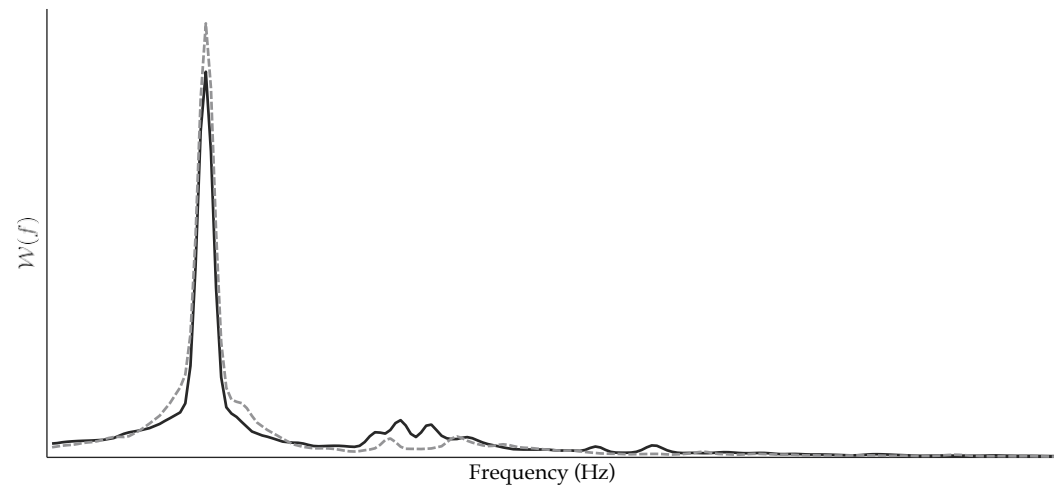

**Figure 3.** Averaged spectral distributions for all the 6 s ECG segments from patients who relapsed to AF (dashed, gray line) and those who maintained SR after the follow-up (solid, black line) once they were aligned with respect to the DF.

To further delve into the analysis of the $f$-wave harmonic spectral structure, four entropy-based indices were computed for the first time from three frequency bands. All intended to quantify how power was distributed along the spectra and then to separately and globally characterize the shape of the DF and its harmonic content. However, each index presented a different mathematical definition and were thus sensitive to different shades in a time series [53]. The Wiener entropy $F$ has been widely studied to predict the results of electrical defibrillation in ventricular fibrillation [37] and is computed by dividing geometric and arithmetic means of the signal spectrum. In a completely different way, spectral entropy $S$ and Rényi spectral entropy $R$ treat the normalized power distribution of the $f$-waves as a probability distribution and calculate Shannon entropy and Rényi entropy, respectively. The first index has already been proposed to discern between patients with persistent and permanent AF [39] and to anticipate spontaneous termination of paroxysmal AF episodes [40]. It should be noted that Rényi entropy is a generalization of Shannon entropy and is thus able to provide additional information [41]. In fact, in the present work, $R$ provided notably better results than $S$ when computed on the LF and TF bands. In the former case, although both indices did not provide statistically significant differences between patients who relapsed to AF versus those who maintained SR after the follow-up, the index $R_{LF}$ reported a nearly significant difference ($p$ = 0.065, see Table 2). Regarding the TF band, $R_{TF}$ provided statistically significant improvements of about 5% in values of Acc, AUC, PPV, and NPV compared to $S_{TF}$ according to McNemar's test. Finally, the $C_0$ complexity divides spectral distribution of a time series to estimate its regular and irregular parts and then obtains an organization estimate highly robust to noise and non-stationary artifacts [42].

Despite slight differences in their results, the indices $F$, $R$, and $C_0$ reported similar general trends on the three analyzed frequency bands of the $f$-waves. In the case of the LF band, negligible differences on the edge of being statistically significant were noticed by the three parameters (see Table 2). This finding suggests that the DF component presented a kindred shape for the patients who relapsed to AF versus those who maintained SR after the follow-up. However, the higher values of entropy and complexity noticed in the patients who maintained SR, along with the trend to higher power peaks ($W(f_0)$) in those who relapsed to AF, point toward the presence of a mildly larger and more-peaked DF component when the probability of AF recurrence after CA increased. This suggestion is supported by Figure 3, where a slightly larger and wider dominant peak can be seen on average for the patients who relapsed to AF. Regarding the HF band, no differences were noticed by any entropy-based parameter, suggesting that both groups of patients presented a similarly well-defined structure of harmonic components of the DF, regardless of the power contained by them. This assumption can also be visually corroborated in Figure 3. Nonetheless, it should be remarked that the indices $F$, $R$, and $C_0$ consider the power distribution along the spectrum, but not the absolute level of power [53], which was highly variable within each group of patients according to the values provided by the variable $W(f_1)$ in Table 2.

These results for the LF and HF bands contrast with those obtained when the TF band, ranging from 3 to 25 Hz, was analyzed. In this case, the parameters $F_{TF}$, $R_{TF}$, and $C_{0TF}$ found highly marked, statistically significant differences and notably high classification performance between the patients who relapsed to AF and those who maintained SR after the follow-up. Indeed, they presented statistically significant improvements of about 10% in values of Se, Sp, Acc, AUC, PPV, and NPV regarding the DF and their computation on the LF band, i.e., $F_{LF}$, $R_{LF}$, and $C_{0LF}$ (see Table 3). Moreover, along with the index $\gamma$, the variable $C_{0TF}$ reported the highest performance metrics, with values about 70%. This wide disparity among frequency bands and the good results of the index $\gamma$ suggest that the most useful information for preoperative prediction of CA outcome lies in the relation between the DF and its harmonic structure and not in the single information individually provided by each one. In this respect, it is interesting to highlight that the parameters $F_{TF}$, $R_{TF}$, and $C_{0TF}$ globally evaluated the whole spectrum of the $f$-waves before taking into consideration both the DF and its harmonic content. The greater entropy and complexity values observed in the patients who maintained SR during the follow-up could hence be explained by the fact that they presented larger harmonics and a proportionally smaller peak DF than those who relapsed to AF. This assumption can be visually corroborated in Figure 3 on average for the two groups of patients.

The same idea of the presence of a larger DF component with low-amplitude harmonics in the patients who relapsed to AF also underlies the values reported by the variable $\gamma$. However, the relation of this index with $F_{TF}$, $R_{TF}$, and $C_{0TF}$ was not as strong as initially expected. In fact, parametric analysis reported notably low correlation values, which were on the edge of being statistically significant, of 16.36% ($p = 0.045$), 11.96% ($p = 0.144$), and 17.96% ($p = 0.027$) for the pairs $\gamma$ and $F_{TF}$, $\gamma$ and $R_{TF}$, and $\gamma$ and $C_{0TF}$, respectively. Moreover, the conducted multivariable analysis also provided that the combination of these pairs of parameters in LDA-based prediction models obtained statistically significant improvements of about 5% in values of Se, Sp, Acc, and AUC regarding the single variables (see Tables 2 and 3), thus achieving the best performance metrics of about 75%. Similarly, PPV and NPV also experienced a statistically significant increase of about 5% regarding the single parameters combined in these models and of up to 20% in comparison with the DF. As a consequence, the index $\gamma$ and the entropy-based parameters computed on the TF band could contain complementary information, and therefore, joint analysis of both the power ratio between the DF and its first harmonic and the global distribution of the power along the spectrum seems to play a key role to improve the preoperative prediction of CA outcome in persistent AF patients.

Extensive ablation based on linear lesions or complex fractionated electrogram (CFE) ablation in addition to PVI has been widely proposed in the cientific literature and widely used in clinical practice. However, large, multicenter and prospective studies comparing these CA strategies with PVI alone have failed to provide any additional benefit, but prolong the duration of the procedure [54,55]. Thereby, all patients enrolled in the present study were treated with PVI alone. The adoption of this CA strategy provides the fairest comparison between patients. In fact, all subjects received the same atrial substrate modifi-cation, thus avoiding a potential bias in the study and allowing the analyzed preoperative predictors to focus on quantifying main differences between atrial electrophysiological features of the patients with high and low risk of midterm AF recurrence. Contrarily, when a tailored CA protocol is applied to each patient by PVI plus a variable number of linear lesions or targeted CFEs, it is reasonable to think that the preoperative predictors could be impacted by the different degree of modification provoked in the atrial substrate of each patient. However, to the best of our knowledge, this hypothesis has still not been corroborated. Some previous works have considered individualized protocols for each patient, but disaggregated data on each kind of intervention (i.e., PVI alone, PVI plus linear lesions, and PVI plus CFE ablation) have not been provided to date [15,34–36,52,56]. These studies have reported a high disparity in the predictive power and cut-off point for some well-established parameters, such as the DF, and the use of different CA protocols could partially explain this finding. Clearly, the proposed predictors based on the $f$-wave harmonic spectral structure could be similarly impacted by the use of tailored CA approaches, but this aspect will have to be addressed in further studies.

Finally, some limitations of the study merit comment. Following current clinical guidelines [7], follow-up of patients after CA was mainly based on standard ECG and 24 h Holter monitoring at different scheduled visits and, in case of symptoms, additional exploration and ECG recordings in the emergency room. However, since continuous ECG monitoring for the whole follow-up was not used, some asymptomatic, non-sustained AF episodes may have been overlooked, and the number of patients with AF recurrence may have been underestimated. Moreover, a follow-up of 9 months could be considered a short period of time to assess CA outcome; however, this midterm AF recurrence prediction is still clinically interesting to select optimal candidates for the intervention. In fact, many previous works have addressed CA outcome predictions at similar or even shorter periods, such as [15–17,34–36]. Nonetheless, a longer follow-up period will be considered in future work, since visits every 12 months are scheduled for all the patients. On the other hand, the sole analysis of the V1 lead precluded specific information from the electrical activity registered at many atrial regions. Although this lead has proved to be the best to reflect global activation of the atria and has, moreover, been widely analyzed in the scientific literature [16,17,23,57], there is recent evidence that the study of spatial variability of ECG-based parameters could be helpful to improve CA outcome prediction [58]. Hence, a multi-lead extension of $f$-wave harmonic structure analysis will be conducted in the future. Lastly, although a comparable [34–36] or much larger number of patients [23,52,56,58] than in previous works also dealing with CA outcome prediction has been analyzed in the present study, the database was retrospective and only came from two centers. Hence, in the future, the conclusions will be corroborated in wider datasets prospectively collected from a higher number of hospitals. A broader analysis will also be performed, considering more predictors, since the results provided by the indices of this study had a limited PPV value in comparison to NPV.

## 5. Conclusions

The present work has conducted a pioneering analysis of the $f$-wave harmonic spectral structure to improve preoperative prediction of CA outcome in persistent AF patients. The results show that the relation between the DF and its harmonic content contains more-relevant information for prediction than separate analysis of each frequency component. While the DF and its harmonic structure individually presented similar global shapes both for patients who relapsed to AF and those who maintained SR after the follow-up, the power ratio between both components had the best discriminant ability. Indeed, the presence of larger harmonics and a proportionally smaller DF peak was strongly associated with a decreased probability of AF recurrence after CA. Moreover, analysis of the global distribution of the $f$-wave power along the spectrum through diverse entropy-based indices, jointly considering both the DF and its harmonic content, also revealed complementary information with respect to their power ratio, thus significantly improving the preoperative prediction of CA outcome.

**Author Contributions:** Conceptualization, P.E., J.J.R. and R.A.; methodology, P.E., J.J.R. and R.A.; software, P.E., J.R. and M.G.; validation, J.J.R. and R.A.; resources, M.A.A., V.M.H. and S.C.; data curation, P.E., J.R., M.G. and R.A.; writing—original draft preparation, P.E.; writing—review and editing, J.R., M.G., M.A.A., V.M.H., S.C., J.J.R. and R.A. All authors have read and agreed to the published version of the manuscript.

**Funding:** This research received financial support from public grants PID2021-00X128525-IV0, PID2021-123804OB-I00, and TED2021-130935B-I00 of the Spanish Government 10.13039/501100011033 jointly with the European Regional Development Fund (EU), SBPLY/17/180501/000411 and SB-PLY/21/180501/000186 from Junta de Comunidades de Castilla-La Mancha, and AICO/2021/286 from Generalitat Valenciana. Moreover, Pilar Escribano holds a predoctoral scholarship 2020-PREDUCLM-15540, which is co-financed by the operating program of the European Social Fund (ESF) 2014–2020 of Castilla-La Mancha.

**Institutional Review Board Statement:** The study was conducted according to the guidelines of the Declaration of Helsinki, complied with Law 14/2007, 3rd of July, on Biomedical Research and other Spanish regulations, and was approved by the Ethical Review Board of University Hospitals of Toledo and Albacete, Spain (protocol code 5064, 1 December 2020).

**Informed Consent Statement:** Informed consent was received from all the subjects participating in the present research. All acquired data were anonymized before processing.

**Data Availability Statement:** The data supporting reported results and presented in this study are available on request from the corresponding author.

**Conflicts of Interest:** The authors declare no conflict of interest.